\documentclass[a4paper,11pt]{article}
\usepackage{pos}

\usepackage{enumitem}
\setlist[enumerate]{itemsep=1mm}

\title{The Southern Wide-field Gamma-ray Observatory: Status and Prospects}
 \ShortTitle{The Southern Wide-field Gamma-ray Observatory}

\author*[a]{Jim Hinton}
\affiliation[a]{Max-Planck-Institut f\"ur Kernphysik, Postfach 103980, D 69029 Heidelberg, Germany}

\forColl{ SWGO }

\emailAdd{swgo\_spokespersons@swgo.org}

\abstract{The Southern Wide-field Gamma-ray Observatory (SWGO)
  Collaboration is currently engaged in design and prototyping work
  towards the realisation of this future gamma-ray facility. SWGO will
  complement CTA and the existing ground-particle based-detectors of
  the Northern Hemisphere (HAWC and LHAASO) with a very wide field and
  high duty cycle view of the southern sky. Here I
 summarise the status of the project and plans for the future,
  including expectations for sensitivity and science targets as well
  as the status of the site search and technological developments.
}

\FullConference{37$^{\rm{th}}$ International Cosmic Ray Conference (ICRC 2021)\\
		July 12th -- 23rd, 2021\\
		Online -- Berlin, Germany}

\begin{document}

\maketitle

%\section{...}

%The Southern Wide-field Gamma-ray Observatory is a project to build a very high - to ultra high energy gamma-ray detector based on ground-level particle detection, high in the Andes. 
%The project is currently in its R\&D phase…

\section{Motivations for a southern wide-field gamma-ray observatory}

Ground-based gamma-ray astronomy is now firmly established as an astronomical discipline, providing broad insights in the domain of high-energy/non-thermal astrophysics and 
also providing opportunities to probe physics beyond the standard model. Two distinct approaches have been developed to detect the cascades of particles initiated by $\sim$$10^{10}$ eV photons: direct detection of shower particles at mountain altitudes; and the imaging of Cherenkov light emitted by shower particles. Powerful arrays of imaging atmospheric Cherenkov telescopes (IACTs) are present in both hemispheres: MAGIC and VERITAS in the North and HESS in the South. In the case of ground-level particle detectors however, only the Northern hemisphere is well-equipped, with the HAWC and LHAASO installations. 

IACTs provide $\sim$0.1$^{\circ}$, 15\% energy resolution and background rejection power of typically $10^{-3}-10^{-2}$ (with $>70$\% gamma-ray efficiency) even at low ($\sim$ $10^{11}$ eV) energies and have been the workhorses of the field (see e.g.~\cite{WHJHrev}) since the technique was established in 1989~\cite{CrabWhipple}. They are limited, however, by modest field of view (FoV; $\sim$4$~^{\circ}$ diameter) and duty cycle ($\sim15$\%), due to the need for darkness and clear skies.

More recently the ground-level particle technique has been established as a powerful complementary approach. Pioneered by MILAGRO~\cite{milagro} and then \cite{HAWC}, the most recent demonstration of the effectiveness of this approach comes from LHAASO~\cite{LHAASOpev}. With coverage of the entire overhead sky ($\sim$steradian FoV) and close to 100\% duty cycle, such instruments can provide all sky mapping and monitoring capabilities, albeit with more modest (in particular energy-) resolution. Whilst background rejection is a challenge for such instruments in the $\sim$$10^{11}$ eV domain, extremely effective rejection ($10^{-4}-10^{-5}$) has been achieved at and beyond $10^{14}$ eV~\cite{HAWCperf, LHAASOperf}, making such detectors an extremely powerful probe of the most extreme cosmic accelerators.

%%%Astro para
The mapping of the Northern Sky with these instruments has been extremely successful, with highlights including the HAWC detections of very extended emission around the Geminga and Monogem pulsar wind nebulae~\cite{Geminga}, and of the jets of SS~433~\cite{HAWC_SS433}. The dramatic recent success of LHAASO in detecting Galactic source up to the PeV range~\cite{LHAASOpev, LHAASOCrab} is built on the unprecedented level of background rejection that could be achieved. The recent detection of emission from GRBs up to TeV energies (see e.g.~\cite{HESSgrb}) is also a strong motivation to aim for all-sky coverage with TeV facilities with very wide-field coverage.

The installation of the first major ground-particle based gamma-ray observatory in the southern hemisphere therefore presents a huge opportunity for high-energy astrophysics and BSM physics. Many key regions and objects are uniquely accessible from a southern site, including of course the Galactic Centre, (most of) the Fermi Bubbles and the inner Galactic plane, but also several key (relatively) local extragalactic objects. A unique opportunity also exists to detect a high mass (thermal-relic) WIMP annihilating in the central parts of our galaxy, even without a strong central cusp in the Dark Matter (DM) density~\cite{AionDM}, ideally complementing the planned studies with CTA~\cite{CTAscience}.

Adding a powerful wide-field TeV instrument to the Southern Hemisphere would represent a major boost to a broad spectrum of non-thermal astrophysics and astroparticle physics.

\section{The SWGO Collaboration}
  
The SWGO collaboration~\cite{swgo} was founded in 2019 around a joint vision for
ground-particle-based gamma-ray observatory high in the Andes. SWGO
was a union of precursor projects including SGSO~\cite{StrawMan} and
LATTES~\cite{lattes}. SWGO is a truly global collaboration with full
partner institutes in Argentina, Brazil, Chile, Czech Republic,
Germany, Italy, Mexico, Peru, Portugal, South Korea, the United
Kingdom, and the United States of America, and supporting scientists
in 10 additional countries. 61 partner institutes have now signed the 'Statement of Interest'
for development of SWGO. The strong scientific involvement of South
American countries is very important to us in our efforts to establish
a new observatory on the continent. SWGO partner institutes bring with
them a rich heritage from many relevant projects including HAWC,
the Pierre Auger Observatory, IceCube, MAGIC and HESS.

\section{R\&D Phase Plan}

SWGO is currently in a research and development phase which ends with the completion of our Conceptual Design Report for SWGO construction and operation. The key activities of this phase are to establish a Baseline design and identify a preferred Site. Table~\ref{tab:milestones} lists the Milestones of the R\&D phase. Milestones M1-3 are complete and M4-5 are anticipated by the end of the year (see Section~\ref{sec:progress}. Completion of the R\&D phase is expected by the end of 2023.

We are performing a science-based optimisation of the observatory in a multi-step process towards site selection, observatory design and the planning of observatory operations. The plan (M1) details the definition of benchmark science cases (M2) to be used to evaluate candidate configurations (M5,M6) and support finalisation of the design (M8). A reference configuration and technology options to evaluate are defined in M3 (see section~\ref{sec:progress}).  The site search process in SWGO is separated in to steps of candidate identification, site short-listing (M4), and finally the identification of a preferred site for the project (M7). Finally the detailed plans for SWGO construction and operation are captured in a Conceptual Design Report (CDR).

The R\&D phase will be followed by a Preparatory Phase for finalisation of the engineering and project management, as well as resource identification. Construction of the array is anticipated to be staged to allow for lessons learned to be incorporated and reduce risks.

%TAB: Milestones 
\begin{table}[h!t]
    \centering
  \begin{tabular}{ll}
\hline\hline
{\bf M1}	& R\&D Phase Plan Established \\
{\bf M2}	& Science Benchmarks Defined \\
{\bf M3}	& Reference Configuration \& Options Defined \\
{\bf M4}	& Site Shortlist Complete \\
{\bf M5}	& Candidate Configurations Defined \\
{\bf M6}	& Performance of Candidate Configurations Evaluated \\
{\bf M7}	& Preferred Site Identified \\
{\bf M8}	& Design Finalised \\
{\bf M9}	& Conceptual Design Report Complete \\
\hline\hline
%Construction \& Operation Proposal Complete \\
\end{tabular}
\caption{The Milestones of the current SWGO research and development phase.}
\label{tab:milestones}
\end{table}

\section{Detector Options and Site Candidates}

\begin{figure}
\includegraphics[width=0.99\textwidth]{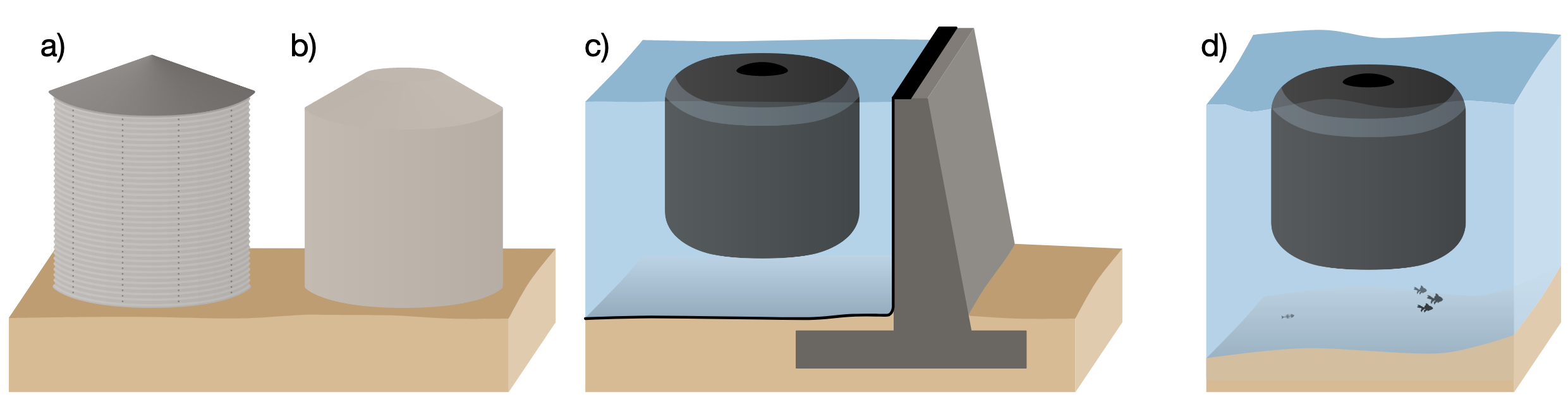}
\caption{Detector concepts under study: cylindrical tanks constructed from (a) corrugated steel sheets or
(b) roto-moulded HDPE; (c) open pond with floating bladder; (d) natural lake with floating bladder (see~\cite{Felix} for more details).}
\label{fig:options}
\end{figure}

The primary detector technology options for the water Cherenkov detector units of SWGO are (see Figure~\ref{fig:options}):
\begin{itemize}
\item {\bf Tanks: } individual detector units mechanically separated and individually deployed, with light tight liners within roto-moulded plastic (as in Auger) or steel (as in HAWC) tanks.
\item {\bf Ponds:} four or more large artificial water volumes with retaining walls and optical separation between units (as in the WCD of LHAASO).
\item {\bf Lake:} deployment of detector unit bladders filled with pure water directly in to a natural lake~\cite{Hazal}.
\end{itemize}
All of these physical unit options as well as photosensor and electronics options are under active evaluation as described in~\cite{Felix}.

Given the otherwise overwhelming background of charged cosmic ray showers, background rejection is a central performance driver for SWGO. A long-established discriminant between gamma and background showers is muon content, and the power of this approach has recently been spectacularly demonstrated by LHAASO~\cite{LHAASOperf}. Two different approaches are being pursued  in SWGO to tag muons passing through individual detector units: a double layer approach~\cite{Kunwar} and multiple-PMTs in a single shallow layer~\cite{Ruben}.

Table~\ref{tab:sites} summarises the site candidates for SWGO, prior to shortlisting. A range of excellent high-alitude sites is available and very detailed characterisation work will be needed to make a decision. At the Peruvian sites of Sibinacocha and Imata natural lake options exist (see below). Tanks can be deployed at all sites with the exception of the Sibinocha area, for more information see~\cite{Doro}.

\begin{table}[h!t]
    \centering
    \begin{tabular}{ll|ccp{3cm}}
\hline\hline
Country & Site Name & Latitude & Altitude & Other installations\\
 & &  & [m a.s.l] & \\
\hline\hline
Argentina & Alto Tocomar & 24.19 S & 4,430 &  \\
         & Cerro Vecar & 24.19 S & 4,800 &  LLAMA, QUBIC \\
         \hline
Bolivia & Chacaltaya & 16.23 S & 4,740 & ALPACA\\
\hline
Chile   & Pajonales & 22.57 S & 4,600 & ALMA and others\\
     & Pampa La Bola & 22.56 S & 4,770 & ALMA and others\\
\hline
Peru & Imata & 15,50 S & 4,450 & \\
     & Yanque & 15.44 S & 4,800 & \\
     & Sibinacocha & 13.51 S & 4,900 &  \\
\hline\hline
    \end{tabular}
    \caption{
      Locations of SWGO candidates sites in South America -- from~\cite{Doro}.
    }
    \label{tab:sites}
\end{table}

\section{Progress so far}
\label{sec:progress}

Since our first collaboration meeting in Padova in October 2019 the collaboration has made significant progress in a number of areas. These are addressed briefly in turn below.

{\bf Definition of Science Benchmarks}. There is no unique performance parameter that can guide the optimisation of the SWGO. For this reason we have defined a number of Benchmarks for SWGO. The benchmarks are designed to be well defined/quantifiable, emphasising a wide range of instrument response characteristics and covering a wide range of key SWGO target science. Table~\ref{tab:benchmarks} lists the defined cases, which will be evaluated for our Candidate Configurations during 2022, and help to inform the major decisions of the project on array layout and detector unit options. See~\cite{Barres}, in these proceedings, for details.
%%%TAB: Science Benchmarks

\begin{table*}
    \centering
    \begin{tabular}{|p{4.5cm}|p{4.5cm}|p{5.cm}|}
        \hline
        %\rowcolor{SWGOOrange}
        \large{\textbf{{Science Case}}} & \large{\textbf{{Design Drivers}}} & \large{\textbf{{Benchmark Description}}} \\
        \hline \hline
        Transient Sources: \newline Gamma-ray Bursts & Low-energy sensitivity \& \newline Site altitude$^a$ & Min. time for 5$\sigma$ detection: \newline F(100~GeV) = $10^{-8}$ erg/cm$^2$.s, \newline PWL index = -2., F(t) $\propto t^{-1.2}$ \\
        \hline
       Galactic Accelerators: \newline PeVatron Sources & High-energy sensitivity \& \newline Energy resolution$^b$ & Maximum exp-cutoff energy detectable 95\% CL in 5 years for: \newline F(1TeV) = 5 mCrab, index = -2.3 \\
        \hline
          Galactic Accelerators: \newline PWNe and TeV Halos & Extended source sensitivity \newline \& Angular resolution$^c$ & Max. angular extension detected at 5$\sigma$ in 5-yr integration for: \newline F(>1TeV) = 5$\times 10^{-13}$ TeV/cm$^{-2}$.s \\
        \hline
          Diffuse Emission: \newline Fermi Bubbles & Background rejection & Minimum diffuse cosmic-ray residual background level. \newline Threshold: < 10$^{-4}$ level at 1 TeV. \\
        \hline
          Fundamental Physics: \newline Dark Matter from GC Halo & Mid-range energy sensitivity \newline Site latitude${^d}$ & Max. energy for $b\bar{b}$ thermal relic cross-section limit at 95\% CL in 5-years, for Einasto profile.\\
        \hline
          Cosmic-rays: \newline Mass-resolved dipole / \newline multipole anisotropy & Muon counting capability$^e$ & Max. dipole energy at 10$^{-3}$ level; Log-mass resolution at 1 PeV -- goal is A=\{1, 4, 14, 56\}; Maximum multipole scale > 0.1 PeV \\
        \hline
    \end{tabular}
    \caption{Science benchmarks defined for SWGO. Instrument response functions derived for candidate SWGO configurations will be used to derive the expected performance in each Benchmark cases, allow a scientificly motivated design decision to be made.}
    \label{tab:benchmarks}
\end{table*}
%Science Case Design Drivers Benchmark Description

%\begin{table}
%  \begin{tabular}{llllll}

%Science Benchmarks defined

{\bf Definition of Reference Configuration}. A starting point for performance evaluation and design development was the complete definition of an array layout and detector unit design that can serve as a reference for future studies. The SWGO Reference Configuration is based on 3.8 m diameter and 3 m deep tanks divided in to a top layer and a muon layer. A dense central array (fill factor $\approx$80\%) extends out to 160~m, with a much lower density (5 \%) outer detector is implemented with the same detector units out to 300~m. For more details see~\cite{Harm}.

{\bf Evaluation of Candidate Sites}. Extensive data collection has taken place for the sites listed in table~\ref{tab:sites}. Considerations include the altitude, local topology, environment, site access/transport costs, and the availability/cost of water, power and network connectivity. An
autonomous station for environmental characterisation~\cite{aerosite} has been developed, and is now being deployed at candidate sites.

{\bf Progress in analysis and simulations}. The simulation and analysis framework of SWGO is built on CORSIKA~\cite{CORSIKA} and the framework developed for HAWC,
%~\cite{HAWCsim},
which in turn makes use of GEANT-4~\cite{geant4} for simulations of individual detector units. A wide range of detector unit options have now been implemented and analysis optimisation is well underway~\cite{Harm}. Highlights include the demonstration of promising gamma/hadron approaches~\cite{Ruben, Kunwar} and the potential for excellent angular resolution for events falling in the central array~\cite{Hofmann, Kunwar}. We are now preparing candidate configurations (M5) for scientific evaluation. The idea is to explore a wide phase space of array layouts and detect unit configurations, to home in on a the most promising region for further optimisation, and tailoring of the design to the chosen site, to fix the design for milestone M8.

\begin{figure}
\includegraphics[width=0.99\textwidth]{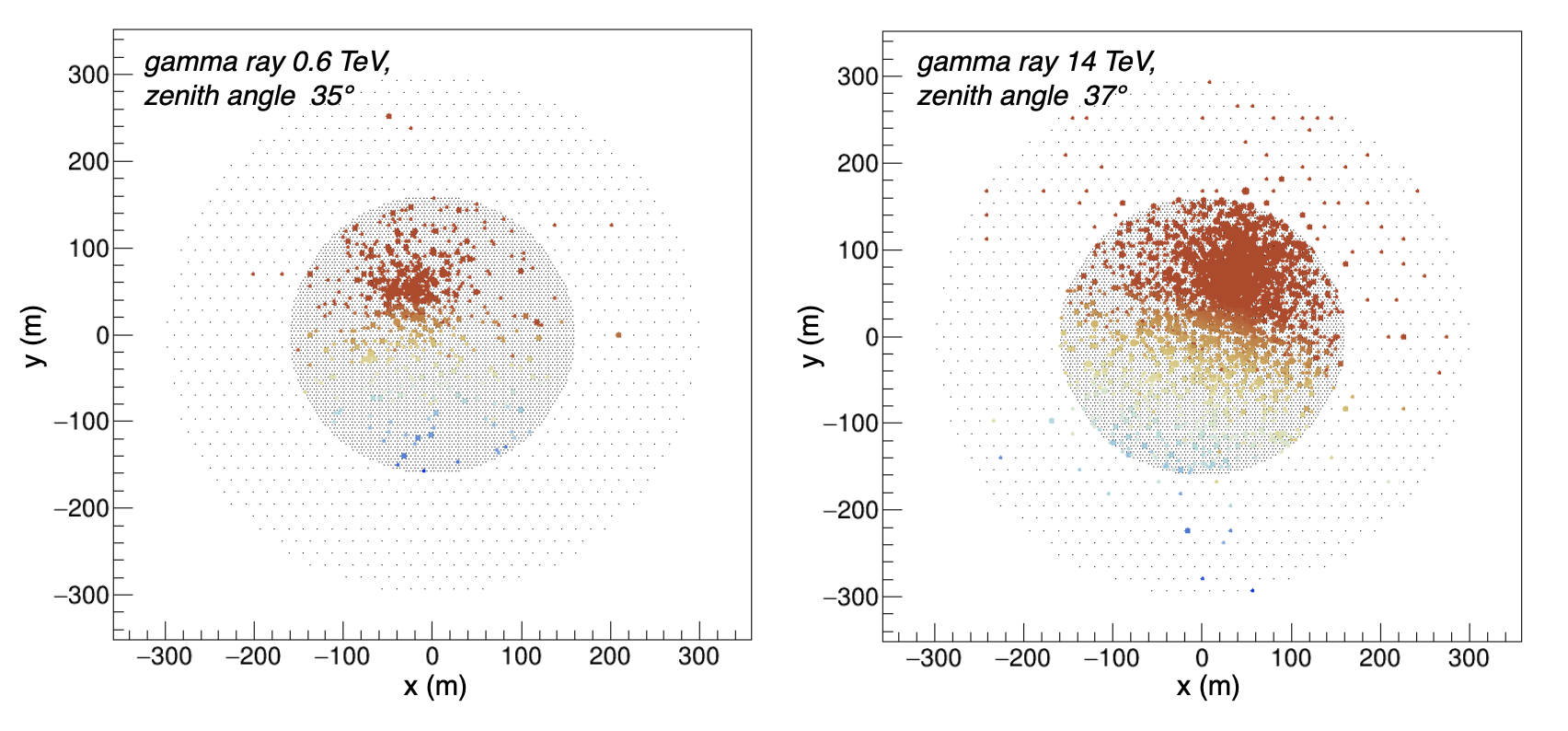}
\caption{Two example $\gamma$-ray events simulated for the Reference Configuration of SWGO. Marker size indicates the amplitude of the measured signal in each detector unit and colour is used to indicate the measured arrival time of the first particle in a unit.}
\label{fig:events}
\end{figure}

Figure~\ref{fig:events} shows two simulated events for the Reference Configuration as examples. Very detailed measurements are possible for showers over a wide range of zenith angles down to a few hundred GeV, at energies above $\sim$10 TeV very precise angular reconstruction is possible for events which fall in the main array.
%% IRFs?

{\bf Detector developments}. A detailed evaluation including technical development working and costing exercises are well underway for the tank, pond and lake options. Tank-based prototypes are being prepared for deployment at candidate sites, based on  both roto-moulded plastic and assembled steel tank approaches. A 10~m diameter, 7~m deep, test pool has been constructed at MPIK for the development of the natural lake concept~\cite{Hazal}, within which a prototype unit is now being tested. In parallel we are developing the concept for DAQ and triggering and evaluating photosensor and electronics options. There is an end-to-end (PMT - software) demonstration of a signal/HV/digitisation concept (using a system developed for CTA~\cite{FC}) which is now undergoing accelerated stress-testing to assess reliability.

{\bf Outreach}. Videos and leaflets have been prepared so far in English and Spanish language, with the intention to expand to additional languages and to reach the communities around the sites with a positive message about the project, aiming to build a strong constructive relationship for the future.

\section{Scientific prospects}
\label{sec:prospects}

The performance expected for SWGO in terms of point-source sensitivity is provided in Fig.~\ref{fig:sens}. As a full evaluation is yet to take place and the design is not fixed we indicate the range which we hope to probe in simulations over the next year. We now consider the 'strawman' curves provided in \cite{StrawMan} to be a lower limit to the performance that can achieved within our nominal cost envelope, due to improvements in event reconstruction and background rejection. We will assess over the next months if coverage of a LHAASO-like km$^{2}$ area with background-free performance can be realised at ultra-high energies, together with excellent low-energy performance. Our benchmarks described above will allow us to study the trade-offs associated to detector unit and array design, but it is already clear that SWGO will be a powerful instrument in a number of key areas:

\begin{itemize}
\item Very extended emission -- including the diffuse galactic emission, the Fermi bubbles and extended halos around PWN and other accelerators. SWGO will detect the emission of the Fermi Bubbles and allow us to discriminate between hadronic and leptonic scenarios. A significant catalog of PWN and halos will be established, with SWGO as a powerful instrument for the discovery of new nearby pulsars. See~\cite{Lopez} for details.
\item Transient phenomena -- in particular the search for prompt-phase TeV emission from GRBs and gravitational wave-emitting mergers. SWGO will have unprecedented capabilities for the detection of prompt phase emission, with an estimated GRB detection rate of $\sim$1 per year~\cite{swgoGRB}.
\item Beyond standard model physics searches -- including for dark matter annihilation and evaporating primordial black holes. Targeting the inner galaxy / Galactic SWGO can reach the expected cross-section for a DM particle which is a thermal relic of the big bang for a wide range of particle mass, annihilation channel and DM-halo profile shape~\cite{AionDM}.
A SWGO search for primordial black holes will reach an order of magnitude deeper than current VHE searches~\cite{swgoPBH}.
\end{itemize}

%%and Fig.~\ref{fig:res} respectively ...

\begin{figure}
\includegraphics[width=0.95\textwidth]{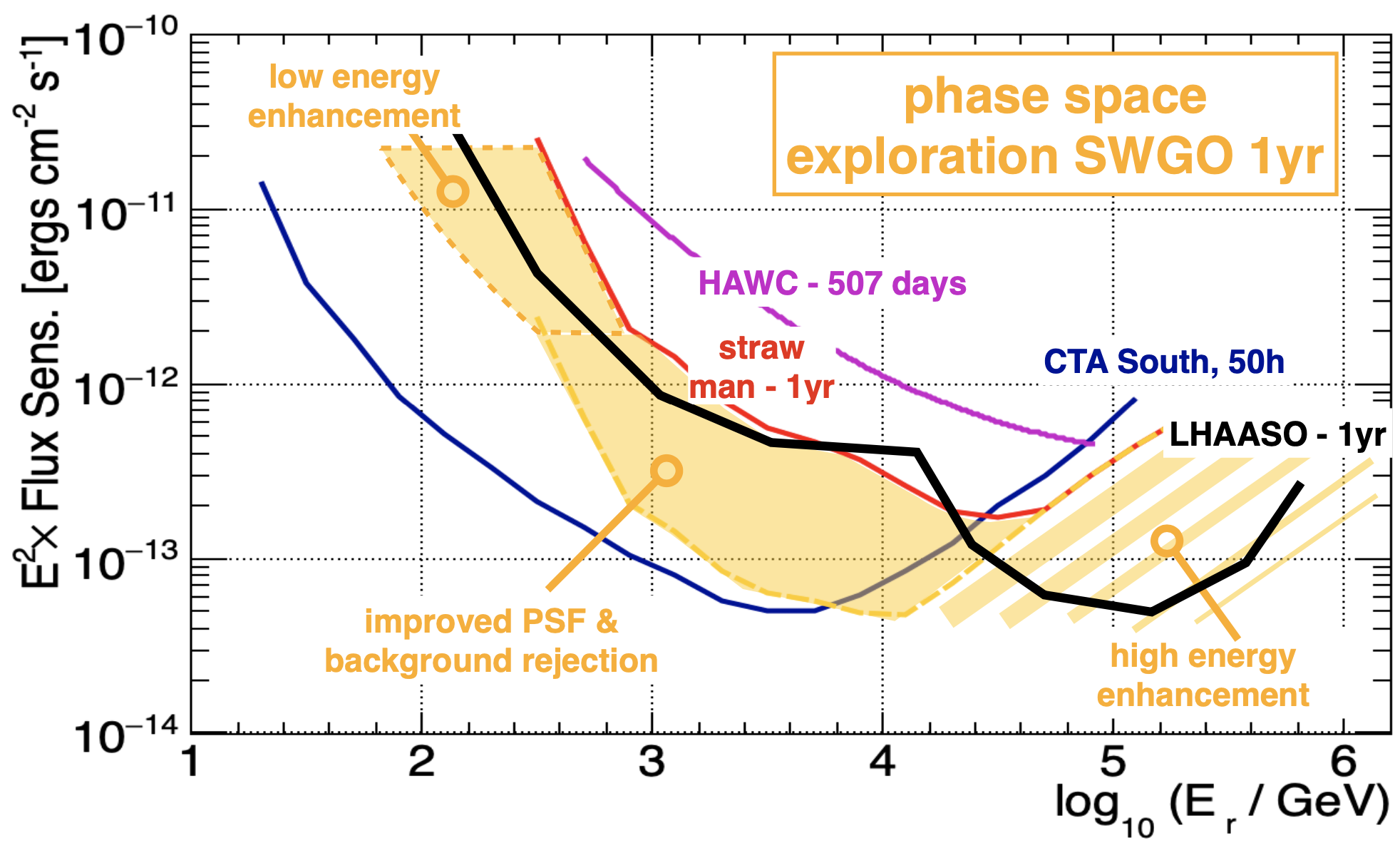}
\caption{Differential point source sensitivity of several experiments\cite{LHAASOSens,CTA,HAWC,StrawMan} and the phase-space that will be explored in the design studies for SWGO (see text for more details).}
\label{fig:sens}
\end{figure}

The combination of SWGO with CTA-South promises to be extremely effective in a number of cases, in particular:
\begin{itemize}
  \item Prompt phase and/or very earlier afterglow emission from GRBs detected using SWGO, with an alert delivered to CTA-S to allow early precision follow-up. SWGO alerts will have much smaller error boxes then (for example) Fermi-GBM and hence of great value to the MWL/MM community.
  \item Triggering of CTA (and the wider community) due to high flux states in AGN and perhaps Galactic variable objects. 
  \item Follow-up with CTA of (in particular) off-Galactic-plane VHE and UHE gamma-ray sources discovered with SWGO. CTA observations will provided excellent angular resolution mapping of such sources.
\end{itemize}

Joint studies with LHAASO are also very promising. The full-sky coverage obtained by combining LHAASO and SWGO observations is important for population studies, for mapping the diffuse emission of the Galaxy and also for cosmic ray anisotropy measurements~\cite{Taylor}.

\begin{figure}
\includegraphics[trim=90 60 90 60, clip, width=0.95\textwidth]{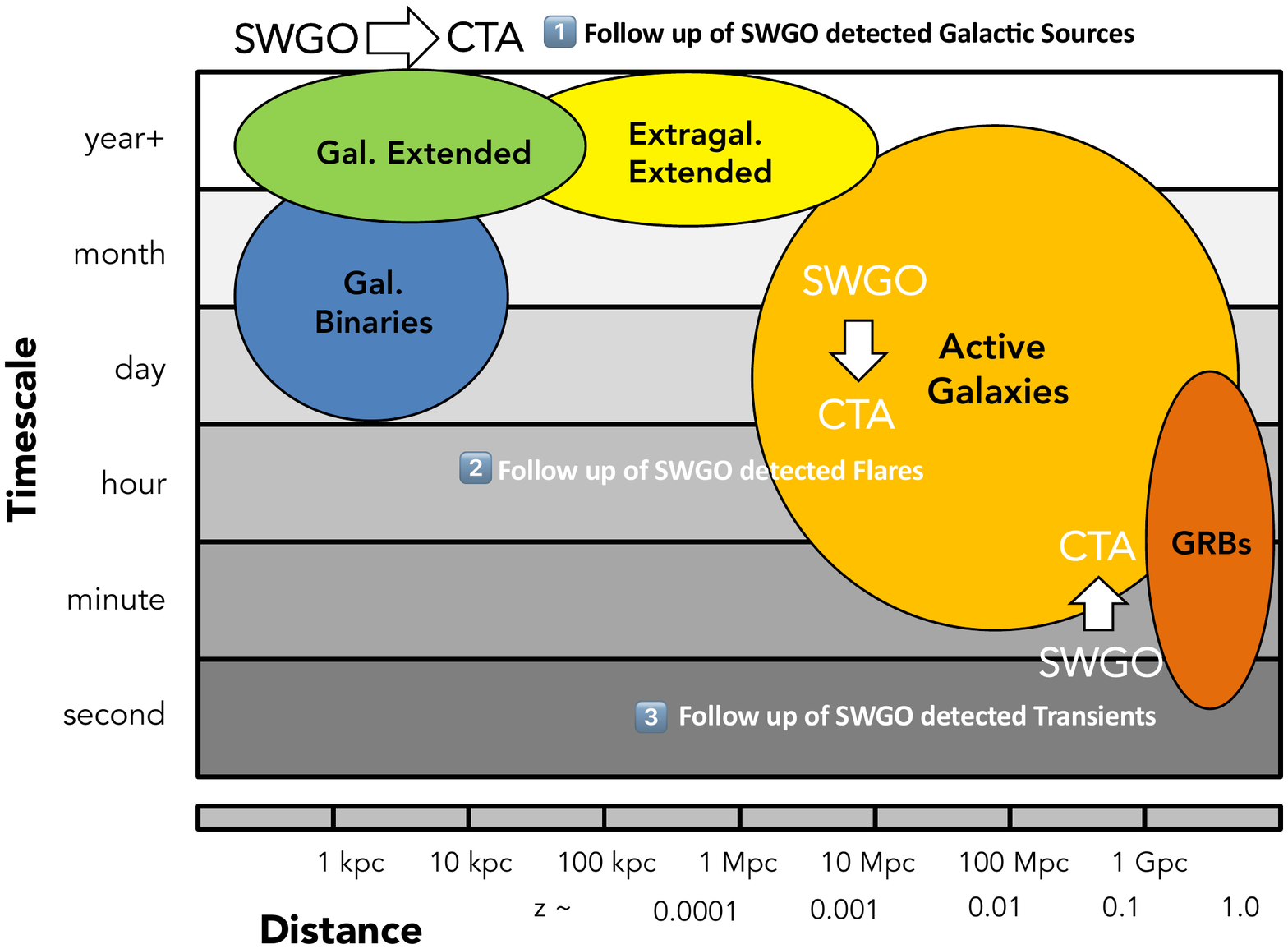}
\caption{Opportunities for SWGO/CTA synergistic operation. Source classes are sorted by their variability time-scales as measured with VHE photons and approximate range of distances. In three cases CTA-South follow-up of SWGO detected sources and/or events is anticipated. See text for details.}
\label{fig:sens}
\end{figure}

\section{Conclusions}

%% \tableofcontents
SWGO is a strongly science-motivated new observatory for the extreme universe. As a powerful wide-field high-duty-cycle observatory in the southern hemisphere it will have very strong synergies with both CTA and LHAASO.
The recent positive appraisal of SWGO in the US Decadal Survey of Astronomy and Astrophysics~\cite{decadal} represents a very welcome boost to the project.
Despite being slowed down significantly by the COVID-19 pandemic, considerable progress has been made towards the SWGO design and we are very optimistic that an instrument with superb performance will be realised in the Andes in the years to come.

\begin{footnotesize}
\paragraph{Acknowledgment} The SWGO Collaboration acknowledges the support from the agencies and organizations listed here: \url{https://www.swgo.org/SWGOWiki/doku.php?id=acknowledgements}.
\end{footnotesize}

%% Full authors list (ONLY FOR COLLABORATIONS)
\clearpage
\section*{Full Authors List: \Coll\ Collaboration}
%
%\noindent \textbf{Note comment afterwards:} Collaborations have the possibility to provide an authors list in xml format which will be used while generating the DOI entries making the full authors list searchable in databases like Inspire HEP. For instructions please go to icrc2021.desy.de/proceedings or contact us under icrc2021proc@desy.de.\\
%
%
\scriptsize
\noindent
P.~Abreu$^1$,
A.~Albert$^2$,
E.\,O.~Angüner$^3$,
C.~Arcaro$^4$,
L.\,H.~Arnaldi$^5$,
J.\,C.~Arteaga-Velázquez$^6$,
P.~Assis$^1$,
A.~Bakalová$^7$,
U.~Barres\,de\,Almeida$^8$,
I.~Batković$^4$
% ...
\vspace{1cm}

\noindent
$^{1}$ {Laboratório de Instrumentação e Física Experimental de Partículas (LIP), Av. Prof. Gama Pinto 2, 1649-003 Lisboa, Portugal\\}
$^{2}$ {Physics Division, Los Alamos National Laboratory, P.O. Box 1663, Los Alamos, NM 87545, United States\\}
$^{3}$ {Aix Marseille Univ, CNRS/IN2P3, CPPM, 163 avenue de Luminy - Case 902, 13288 Marseille cedex 09, France\\}
$^{4}$ {University of Padova, Department of Physics and Astronomy \& INFN Padova, Via Marzolo 8 - 35131 Padova, Italy\\}
$^{5}$ {Centro Atómico Bariloche, Comisión Nacional de Energía Atómica, S. C. de Bariloche (8400), RN, Argentina\\}
$^{6}$ {Universidad Michoacana de San Nicolás de Hidalgo, Calle de Santiago Tapia 403, Centro, 58000 Morelia, Mich., México\\}
$^{7}$ {FZU, Institute of Physics of the Czech Academy of Sciences, Na Slovance 1999/2, 182 00 Praha 8, Czech Republic \\}
$^{8}$ {Centro Brasileiro de Pesquisas Físicas, R. Dr. Xavier Sigaud, 150 - Rio de Janeiro - RJ, 22290-180, Brazil\\}
$^{9}$ {Academic Department of Physics – Faculty of Sciences – Universidad Nacional de San Antonio Abad del Cusco (UNSAAC), Av. de la Cultura, 733, Pabellón C-358, Cusco, Peru\\}
$^{10}$ {Instituto de Física, Universidad Nacional Autónoma de México, Sendero Bicipuma, C.U., Coyoacán, 04510 Ciudad de México, CDMX, México \\}
$^{11}$ {Dipartimento di Fisica, Università degli Studi di Torino, Via Pietro Giuria 1, 10125, Torino, Italy\\}
$^{12}$ {RWTH Aachen University, Physics Institute 3, Otto-Blumenthal-Straße, 52074 Aachen, Germany \\}
$^{13}$ {University of Maryland, College Park, MD 20742, United States\\}
$^{14}$ {Durham University, Stockton Road, Durham, DH1 3LE, United Kingdom\\}
$^{15}$ {Astronomical Observatory, University of Warsaw, Aleje Ujazdowskie 4, 00478 Warsaw, Poland\\}
$^{16}$ {Facultad de Ciencias en Física y Matemáticas UNACH, Boulevard Belisario Domínguez, Km. 1081, Sin Número, Terán, Tuxtla Gutiérrez, Chiapas, México\\}
$^{17}$ {Facultad de Ciencias Exactas, Universidad Nacional de Salta, Avda. Bolivia N° 5150, (4400) Salta Capital, Argentina\\}
$^{18}$ {Department of Aerospace Science and Technology, Politecnico di Milano, Via Privata Giuseppe La Masa, 34, 20156 Milano MI, Italy\\}
$^{19}$ {Facultad de Ciencias Físico Matemáticas, Benemérita Universidad Autónoma de Puebla, C.P. 72592, México\\}
$^{20}$ {Instituto de Astronomia y Fisica del Espacio (IAFE, CONICET-UBA), Casilla de Correo 67 - Suc. 28 (C1428ZAA), Ciudad Autónoma de Buenos Aires, Argentina\\}
$^{21}$ {Universidad de Guadalajara, Blvd. Gral. Marcelino García Barragán 1421, Olímpica, 44430 Guadalajara, Jal., México\\}
$^{22}$ {Instituto de Física de São Carlos, Universidade de São Paulo, Avenida Trabalhador São-carlense, nº 400, Parque Arnold Schimidt - CEP 13566-590, São Carlos - São Paulo - Brasil\\}
$^{23}$ {INFN - Roma Tor Vergata and INAF-IAPS, Via del Fosso del Cavaliere, 100, 00133 Roma RM, Italy\\}
$^{24}$ {Dept. of Physics and CCTVal, Universidad Tecnica Federico Santa Maria, Avenida España 1680, Valparaíso, Chile\\}
$^{25}$ {Universität Würzburg, Institut für Theoretische Physik und Astrophysik, Emil-Fischer-Str. 31, 97074 Würzburg, Germany\\}
$^{26}$ {Department of Physics, and the Wisconsin IceCube Particle Astrophysics Center (WIPAC), University of Wisconsin, 222 West Washington Ave., Suite 500, Madison, WI 53703, United States\\}
$^{27}$ {TUBITAK Space Technologies Research Institute, ODTU Campus, 06800, Ankara, Turkey\\}
$^{28}$ {Friedrich-Alexander-Universität Erlangen-Nürnberg, Erlangen Centre for Astroparticle Physics, Erwin-Rommel-Str. 1, D 91058 Erlangen, Germany\\}
$^{29}$ {Tecnologico de Monterrey, Escuela de Ingeniería y Ciencias, Ave. Eugenio Garza Sada 2501, Monterrey, N.L., 64849, México\\}
$^{30}$ {Max-Planck-Institut f\"ur Kernphysik, P.O. Box 103980, D 69029 Heidelberg, Germany\\}
$^{31}$ {Università di Napoli “Federico II”, Dipartimento di Fisica “Ettore Pancini”,  and INFN Napoli, Complesso Universitario di Monte Sant'Angelo - Via Cinthia, 21 - 80126 - Napoli, Italy \\}
$^{32}$ {University of Granada, Campus Universitario de Cartuja, Calle Prof. Vicente Callao, 3, 18011 Granada, Spain\\}
$^{33}$ {IFLP, Universidad Nacional de La Plata and CONICET, Diagonal 113, Casco Urbano, B1900 La Plata, Provincia de Buenos Aires, Argentina\\}
$^{34}$ {University of Utah, 201 Presidents' Cir, Salt Lake City, UT 84112, United States\\}
$^{35}$ {Michigan Technological University, 1400 Townsend Drive, Houghton, MI 49931, United States\\}
$^{36}$ {Dipartimento di Fisica e Astronomia "E. Majorana", Catania University and INFN, Catania, Italy\\}
$^{37}$ {APC--IN2P3/CNRS, Université de Paris, Bâtiment Condorcet, 10 rue A.Domon et Léonie Duquet, 75205 PARIS CEDEX 13, France\\}
$^{38}$ {University of Leicester, University Road, Leicester LE1 7RH, United Kingdom\\}
$^{39}$ {Department of Physics, University of Trieste and INFN Trieste, via Valerio 2, I-34127, Trieste, Italy \\}
$^{40}${Centro de Investigación en Computación, Instituto Politécnico Nacional, Av. Juan de Dios Bátiz S/N, Nueva Industrial Vallejo, Gustavo A. Madero, 07738 Ciudad de México, CDMX, México\\}
$^{41}$ {Department of Physics and Mathematics, Universidad de Monterrey, Av. Morones Prieto 4500, San Pedro Garza García 66238, N.L., México\\}
$^{42}$ {Department of Physics and Astronomy, University of Alabama, Gallalee Hall, Tuscaloosa, AL 35401, United States\\}
$^{43}$ {Instituto de Tecnologías en Detección y Astropartículas (CNEA-CONICET-UNSAM), Av. Gral Paz 1499 - San Martín - Pcia. de Buenos Aires, Argentina\\}
$^{44}$ {Department of Physics, ETH Zurich, CH-8093 Zurich, Switzerland\\}
$^{45}$ {Instituto de Ciencias Nucleares, Universidad Nacional Autónoma de México (ICN-UNAM), Cto. Exterior S/N, C.U., Coyoacán, 04510 Ciudad de México, CDMX, México\\}
$^{46}$ {Departamento de Física, Facultad de Ciencias Básicas, Universidad Metropolitana de Ciencias de la Educación, Av. José Pedro Alessandri 774, Ñuñoa, Santiago, Chile\\}
$^{47}$ {Department of Physics, University of Seoul, 163 Seoulsiripdaero, Dongdaemun-gu, Seoul 02504, Republic of Korea \\}
$^{48}$ {Institut de recherche sur les lois fondamentales de l'Univers (IRFU), CEA, Université Paris-Saclay, F-91191 Gif-sur-Yvette, France\\}
$^{49}$ {Laboratoire Univers et Particules de Montpellier, CNRS,  Université de Montpelleir, F-34090 Montpellier, France\\}
$^{50}$ {Instituto Nacional de Astrofísica, Óptica y Electrónica (INAOE), Luis Enrique Erro 1, Puebla, México\\}

\end{document}